\documentclass[12pt]{article}
\usepackage{graphicx}
\usepackage[greek,english] {babel}
\usepackage[utf8x] {inputenc}
\usepackage[version=3] {mhchem}
\usepackage {amsfonts}
\usepackage{array}
\usepackage{amsmath}
\usepackage[dvipsnames]{xcolor}
\usepackage{young}
\usepackage[export]{adjustbox}
\usepackage{afterpage}
\usepackage{easybmat}
\usepackage{physics}
\usepackage{authblk}
\usepackage{hhline}
\usepackage{geometry}
\usepackage{times}
\usepackage{nopageno}

\graphicspath{ {c:/Figures/} }

\newcounter{mycite}
\newtoks\citetoks
\makeatletter
\DeclareRobustCommand\unscite[1]{%
  \@ifundefined{uns@cite#1}
    {\refstepcounter{mycite}\label{citelabel@#1}%
     \expandafter\xdef\csname uns@cite#1\endcsname{\arabic{mycite}}%
     \toks\z@=\expandafter{\the\citetoks}%
     \toks\tw@=\expandafter\expandafter\expandafter{%
       \csname uns@bibitem#1\endcsname}%
     \edef\@tempcite{\the\toks\z@\the\toks\tw@}%
     \global\citetoks=\expandafter{\@tempcite}%
    }{}[\@nameuse{uns@cite#1}]}
\newcommand{\mybibitem}[2]{%
  \@namedef{uns@bibitem#1}{\bibitem[\ref{citelabel@#1}]{#1}#2}}
\makeatother

\mybibitem{Rowe}{D. J. Rowe, {\it Rep. Prog. Phys.} 48, 1419 (1985).}
\mybibitem{review}{P. Van Isacker and S. PIttel, {\it Physics Scripta, Vol. 91, Number 2}, (2016).}
\mybibitem{SymplecticI}{G. Rosensteel and D.J. Rowe, {\it Annals of Physics {\bf 123}, 36-60},(1979).}
\mybibitem{SymplecticII}{D. J. Rowe and G. Rosensteel, {\it Annals of Physics {\bf 126}, 198-233}, (1980).}
\mybibitem{SymplecticIII}{G. Rosensteel and D.J. Rowe, {\it Annals of Physics {\bf 126}, 343-370},(1980).}
\mybibitem{Park}{P. Park, J. Carvalho, M. Vassanji and D.J. Rowe, {\it Nuclear Physics {\bf A414}, 93-112}, (1983). }
\mybibitem{Sofia}{D. Bonatsos, S. Karampagia, R. B. Cakirli, R. F. Casten, K. Blaum and Amon Susam, {\it Phys. Rev. C 88, 054309,} (2013).}

\mybibitem{calculation}{D. Bonatsos et al., {\it Phys. Rev. C} {\bf 95}, 064325 (2017). }

\mybibitem{lambda-mu}{ J. P. Draayer, Y. Leschber, S. C. Park and R. Lopez, {\it Comput. Phys. Commun. 56, 279} (1989).}
\mybibitem{Nilssonbook}{S.G. Nilsson and I. Ragnarsson, {\it Shapes and Shells in Nuclear Structure} Cambridge University Press, (1995).}
\mybibitem{dSU(3)}{D. Bonatsos, C. Daskaloyannis, P. Kolokotronis and D. Lenis, {\it arXiv: hep-th/9411218v1}, (1994).}
\mybibitem{wf}{V. X. Genest, L. Vinet, A. Zhedanov, {\it arXiv: 1307.0692v2, math-ph}, (2013). }
\mybibitem{h.w.}{D. Bonatsos, R.F. Casten, A. Martinou, I. E. Assimakis, N. Minkov, S. Sarantopoulou, R. B. Cakirli and K. Blaum, {\it arXiv: 1712.04126v1, nucl-th,} (2017).}
\mybibitem{beta}{O.Castanos, J.P. Draayer, and Y. Leschber, {\it Z. Phys. A329, 33}, (1988).}

\mybibitem{Harvey}  {M. Harvey et al., in {\it Advances in Nuclear Physics}, Vol. 1, (Plenum, New York, 1968) 67. }

\mybibitem{Cohen}{C. Cohen-Tannoudji, B. Diu, F. Laloe, {\it Quantum Mechanics, Volume I} (Herman, Paris, 1977).}
\mybibitem{1Nilsson}{S. G. Nilsson,  Dan Mat Fys Medd 29, no 16 (1955).}
\mybibitem{Bahri} {C. Bahri, J. Escher, J. P. Draayer, Nuclear Physics A 592 (1995) 171-193.}
\mybibitem{Jutta} {J. Escher, C. Bahri, D. Troltenier, J. P. Draayer, Nuclear Physics A 633 (1998) 662-680.}
\mybibitem{64}{J.M. Yao, Z.X. Li, J. Xiang, H. Mei, {\it arXiv: 1010.4372v1}, (2010).}
\mybibitem{QQ}{J. P. Draayer, K.J. Weeks and K. T. Hecht, {\it Nuclear Physics A381, 1-12} (1982).}
\mybibitem{Spheroid}{Wikipedia: Spheroid.}
\mybibitem{Evolution}{R. G. Casten Lectures at Riken, January 2010.}
\mybibitem{Magic}{O. Sorlin, M.G. Porquet, {\it Nuclear Magic Numbers: new features far from stability,} arXiv:0805.2561, (2008). }
\mybibitem{Casten} {R.F. Casten, {\it Nuclear Structure from a Simple Perspective}, (Oxford University Press Inc., New York, 1990).}
\mybibitem{Nuclear-Physics} {W.N. Cottingham and D. A. Greenwood, {\it \textgreek{Εισαγωγή στην Πυρηνική Φυσική}}, (\textgreek{τυπωθήτω-Γιώργος Δάρδανος, Αθήνα 1995}).}
\mybibitem{Greiner}{ W. Greiner, J. A. Maruhn, {\it Nuclear Models}, (Springer, Berlin and Heidelberg, 1996).}
\mybibitem{Casten} {R.F. Casten, {\it Nuclear Structure from a Simple Perspective}, (Oxford University Press Inc., New York, 1990).}

\mybibitem{Bible} {R. Casten et al., {\it Algebraic Approaches to Nuclear Structure}, (Harwood, 1993). }

\mybibitem{W}{ A. R. Edmonds, {\it Angular Momentum in Quantum Mechanics,} (Princeton University Press, Princeton, 1957).}
\mybibitem{Lipkin} {H. J. Lipkin, {\it Lie Groups for Pedestrians} (North-Holland, Amsterdam, 1965).}
\mybibitem{Nuclear-Physics} {W.N. Cottingham and D. A. Greenwood, {\it \textgreek{Εισαγωγή στην Πυρηνική Φυσική}}, (\textgreek{τυπωθήτω-Γιώργος Δάρδανος, Αθήνα 1995}).}

\mybibitem{book} {S. G. Nilsson and I. Ragnarsson, {\it Shapes and Shells in Nuclear Structure} (Cambridge University Press, Cambridge, 1995).}
\mybibitem{Berkeley}{Robert Littlejohn, Lecture notes on Quantum Mechanics, University of Berkeley, (2010). }
\mybibitem{IBM}{F. Iachello, A. Arima, {\it The Interacting boson model}, (Cambridge University Press, 1987).}
\mybibitem{applet}{ http://mathinsight.org/determinant\_linear\_transformation.}
\mybibitem{Wigner-Eckart}{ J. P. Draayer and Yoshimi Akiyama, {\it J. of Math. Phys.,  Vol 14,  No 12}, (1973).}
\mybibitem{rigid-rotator}{ O. Casta$\tilde n$os, J. P. Draayer and Y. Leschber, {\it Z. Phys. A - Atomic Nuclei 329, 33-43,} (1988).}
\mybibitem{Naqvi} {H. A. Naqvi and J. P. Draayer, {\it Nuclear Physics A516, 351-364,} (1990).}
\mybibitem{Gelfand}{ Jes$\acute{u}$s A. De Loera and Tyrrell B. McAllister, {\it Discrete Comput. Geom. 32: 459-470,} (2004).}
\mybibitem{lambda-mu}{ J. P. Draayer, Y. Leschber, S. C. Park and R. Lopez, {\it Comput. Phys. Commun. 56, 279} (1989).}

\mybibitem{Wood}{J.L. Wood et al., {\it Phys.  Rep. {\bf 215}, 101} (1992).}
\mybibitem{coexistence}{ K. Heyde and J. L. Wood, {\it Rev. Mod. Phys. {\bf 83}}, 1467 (2011).}

\mybibitem{CG-coefficients} {C. Bahri, D. J. Rowe, J. P. Draayer, {\it Computer Physics Communications 159, 121-143}, (2004).}
\mybibitem{Elliotta}{ J. P. Elliott, {\it Proc. Roy. Soc. Ser. A 245, 128}, (1958).}
\mybibitem{Elliottb}{ J. P. Elliott, {\it Proc. Roy. Soc. Ser. A Vol. 245, 562}, (1958).}
\mybibitem{Elliottc}{ J. P. Elliott and M. Harvey, {\it Proc. Roy. Soc. Ser. A 272, 557}, (1963).}
\mybibitem{Q} {J.P. Draayer and K. J. Weeks, {\it Annals of Physics, 156, 41-67}, (1984).}
\mybibitem{TE2}{ O. Casta$\tilde n$os, J. P. Draayer and Y. Leschber, {\it Annals of Physics 180, 290-329}, (1987).}
\mybibitem{ENSDF}{ Brookhaven National Laboratory ENSDF database, http://www.nndc.bnl.gov/ensdf/}
\mybibitem{W.U.}{ S. Raman, C.W. Nestor, Jr., and P. Tikkanen, {\it  At. Data Nucl. Data Tables 78, 1},
(2001).}
\mybibitem{Sofia}{D. Bonatsos, S. Karampagia, R. B. Cakirli, R. F. Casten, K. Blaum and Amon Susam, {\it Phys. Rev. C 88, 054309,} (2013).}
\mybibitem{pn pairs} {Dennis Bonatsos, I.E. Assimakis and A. Martinou, {\it Bulg. J. Phys. 42, 439-449}, (2015).}
\mybibitem{Rila} {Andriana Martinou, I.E. Assimakis, N. Minkov, D. Bonatsos, {\it, Nuclear Theory '35, Proceedings of the 35th  International Workshop on Nuclear Theory (Rila 2016),  224-235}, (Heron Press, Sofia, 2016).}

\mybibitem{RilaSmaragda}{S.  Sarantopoulou, A.Martinou, I. E. Assimakis, N. Minkov, and D. Bonatsos, {\it , Nuclear Theory '35, Proceedings of the 35th  International Workshop on Nuclear Theory (Rila 2016), 236-243,} (Heron Press, Sofia, 2016).}

\mybibitem{po} {D. Bonatsos et al., {\it Phys. Rev. C} {\bf 95}, 064326, (2017).}

\mybibitem{Nilsson}{ Dennis Bonatsos, I. E. Assimakis, N. Minkov, Andriana Martinou, R. B. Cakirli, R. F. Casten and K. Blaum, {\it Physical Review C 95, 064325}, (2017).}

\mybibitem{SofiaAndriana}{A. Martinou, D. Bonatsos, I. E. Assimakis, N. Minkov, S. Sarantopoulou, R. B. Cakirli, R. F. Casten, and K. Blaum,{\it  Bulg. J. Phys.  (2017) in press.  Proceedings of the Workshop on ``Shapes and Dynamics of Atomic Nuclei: Contemporary Aspects’’ (SDANCA17, Sofia 2017), ed. N. Minkov}}
\mybibitem{SofiaBonatsos}{D. Bonatsos, I. E. Assimakis, N. Minkov, A. Martinou, S. K. Peroulis, S. Sarantopoulou, R. B. Cakirli, R. F. Casten, and K. Blaum, {\it Bulg. J. Phys.  (2017) in press.  Proceedings of the Workshop on ``Shapes and Dynamics of Atomic Nuclei: Contemporary Aspects’’ (SDANCA17, Sofia 2017), ed. N. Minkov}}

\mybibitem{SofiaSmaragda}{S. Sarantopoulou, D. Bonatsos, I. E. Assimakis, N. Minkov, A. Martinou, R. B. Cakirli, R. F. Casten, and K. Blaum, {\it Bulg. J. Phys.  (2017) in press.  Proceedings of the Workshop on ``Shapes and Dynamics of Atomic Nuclei: Contemporary Aspects’’ (SDANCA17, Sofia 2017), ed. N. Minkov}}

\mybibitem{SofiaGiannis}{I. E. Assimakis, D. Bonatsos, N. Minkov, A. Martinou,  R. B. Cakirli, R. F. Casten, and K. Blaum, {\it Bulg. J. Phys.  (2017) in press.  Proceedings of the Workshop on ``Shapes and Dynamics of Atomic Nuclei: Contemporary Aspects’’ (SDANCA17, Sofia 2017), ed. N. Minkov.}}

\mybibitem{Kr}{ F. Flavigny et al., {\it PRL 118, 242501}, (2017).}
\mybibitem{irreps} {P. Van Isacker, D. D. Warner and D. S. Brenner, {\it Physical review Letters Vol. 74, No 23}, (June 1995).}
\mybibitem{182Hg} {N. Bree et al. {\it Phys. Rev. Lett. 112, 162701}, (April 2014).}
\mybibitem{moments}{T.J. Mertzimekis, K. Stamou, A. Psaltis, Nucl. Instrum. Meth. Phys. Res. A 807, 56 (2016).}
\mybibitem{Ge}{Sugawara, M., Toh, Y., Czosnyka, T. et al. Eur Phys J A (2003) 16: 409.}
\mybibitem{Clement} {E. Cl\'ement et al., {\it Phys. Rev. C 75, 054313,} (2017).}
\mybibitem{70Se}{ A. M. Hurst et al., {\it Phys. Rev. Lett. 98, 072501}, (2007).}
\mybibitem{Asherova}{R. M. Asherova, Yu. F. Smirnov, V. N. Tolstoy, {\it Nuclear Physics, A355, 25-44}, (1981).}

\providecommand{\keywords}[1]
{\small
\textbf{\textit{Keywords}} #1}
\begin{document}
\selectlanguage{english}

\title{\bf Nucleon numbers for nuclei with shape coexistence}

\author[1]{\underline{Andriana Martinou}}
\author[1]{Dennis Bonatsos}
\author[2]{N. Minkov}
\author[3]{T. Mertzimekis}
\author[1]{I.E. Assimakis}
\author[3]{S. Peroulis}
\author[1]{S. Sarantopoulou}
\affil[1]{\footnotesize Institute of Nuclear and Particle Physics, National Centre for Scientific Research  ``Demokritos'', GR-15310 Aghia Paraskevi, Attiki, Greece}
\affil[2]{Institute of Nuclear Research and Nuclear Energy, Bulgarian Academy of Sciences, 72 Tzarigrad Road, 1784 Sofia, Bulgaria}
\affil[3]{University of Athens, Faculty of Physics, Zografou Campus, GR-15784 Athens, Greece}

\maketitle
\begin{center}
\rule{16.5cm}{0.3mm}
\begin{abstract}
\small
We consider two competing sets of nuclear magic numbers, namely the harmonic oscillator (HO) set (2, 8, 20, 40, 70, 112, 168, 240,...) and the set corresponding to the proxy-SU(3) scheme, possessing shells 0-2, 2-4, 6-12, 14-26, 28-48, 50-80, 82-124, 126-182, 184-256... The two sets provide $0^+$ bands with different deformation
and bandhead energies. We show that for proton (neutron) numbers starting from the 
regions where the quadrupole-quadrupole ($Q\cdot Q$) interaction, as derived by the HO, becomes weaker than the one obtained in the proxy-SU(3) scheme, to the regions of HO shell closure, the shape coexistence 
phenomenon may emerge. Our analysis suggests that the possibility for appearance of shape coexistence has to be investigated in the following regions of proton (neutron) numbers: 8, 18-20, 34-40, 60-70, 96-112, 146-168, 210-240,...\\
\flushleft{\keywords{shape coexistence, proxy-SU(3)}}
\end{abstract}

\rule{16.5cm}{0.3mm}
\end{center}
\section{\normalsize INTRODUCTION}
Shell model theory has been established in the nuclear physics community as the major microscopic model for nuclei. In this model the single particle orbitals are organised, so as to have definite angular momentum 
$l$. The algebra of angular momentum is SU(2), with multiplets $\ket{l,l_0}$, with $l_0$ being the projection of angular momentum. But this organisation is sufficient for almost spherical nuclei. In deformed nuclei the previous classification of single particle orbitals is inadequate. The quadrupole interaction prevails and the proper organisation has again an SU(2) structure, but with definite number of quanta in the xy plane $n_\perp=n_x+n_y$, instead of $l$ \unscite{Elliottb}. This SU(2) structure has multiplets $\ket{n_\perp, l_0}$. Two sets of magic numbers, i.e. 20-40 and 28-48, have exactly the same $\ket{n_\perp, l_0}$ multiplets and identical magnitude of the $m=2$ component of quadrupole interaction. Thus the two sets compete.

\section{\normalsize THE SU(3) GLOSSARY}\label{glossary}

The three-dimensional (3D) isotropic harmonic oscillator (HO) Hamiltonian is
\begin{equation}
H={1\over 2m}(p_x^2+p_y^2+p_z^2)+{m\omega ^2\over 2}(x^2+y^2+z^2).
\end{equation}
In the second quantization language one can define bosonic annihilation and creation operators 
\begin{eqnarray}
a_j=\sqrt{m\omega \over 2\hbar}x_j+{i\over \sqrt{2m\omega \hbar}}p_j, \qquad\qquad 
a_j^\dagger=\sqrt{m\omega \over 2\hbar}x_j-{i\over \sqrt{2m\omega \hbar}}p_j,\qquad\qquad  j=x,y,z.
\end{eqnarray}
Obviously the eigenfunctions of the Hamiltonian are $\ket{n_z,n_x,n_y}$ in the Cartesian notation, where $n_z,n_x,n_y$ are the number of quanta in each Cartesian axis. The action of the $a, a^\dagger$ operators is
\begin{eqnarray}
a_x\ket{n_x}=\sqrt{n_x}\ket{n_x-1},\qquad \qquad
a_x^\dagger\ket{n_x}=\sqrt{n_x+1}\ket{n_x+1}.
\end{eqnarray}

Now the problem can be turned into cylindrical coordinates, which are more suitable for deformed nuclei. In order to distinguish the $x$-$y$ plane from the $z$ axis, the left and right quanta operators are being defined \unscite{Cohen}
\begin{eqnarray}\label{a}
a_{\pm}={a_x\mp i a_y\over \sqrt{2}},\qquad \qquad 
a_{\pm}^\dagger={a_x^\dagger\pm i a_y^\dagger\over \sqrt{2}}.
\end{eqnarray}
The operators $a_+$, $a_+^\dagger$ are called right hand operators, because they seem to destroy/create a circular quantum, which rotates like a right hand in the $x$-$y$ plane. For the same reason the other two operators are called left hand operators. The operators in the $z$ axis are renamed $a_0=a_z$, $a_0^\dagger=a_z^\dagger$.

With these ingredients one can define eight operators which are the generators of SU(3) \unscite{Lipkin}
\begin{eqnarray}\label{g}
l_0=a_+^\dagger a_+-a_-^\dagger a_-, \qquad \qquad
l_\pm=\pm\sqrt{2}(a_0^\dagger a_\mp-a_\mp^\dagger a_0),\\
q_{\pm 2}=-\sqrt{6}a_\pm^\dagger a_\mp,\qquad \qquad
q_{\pm 1}=\mp\sqrt{3}(a_0^\dagger a_\mp+a_\pm ^\dagger a_0),\qquad \qquad
q_0=2a_0^\dagger a_0-a_+^\dagger a_+-a_-^\dagger a_-.\label{g2}
\end{eqnarray}
The first three operators ($l_0, l_\pm$) form the algebra of angular momentum, while the last five are the five components of the quadrupole moment. The $l$, $q$ are single particle operators, while the $L$, $Q$ are used for the whole shell and they are derived from the sum over all particles.

The Hamiltonian of the many nucleon problem is
\begin{equation}\label{H}
H=\sum_kH_k-{\chi\over 2} Q\cdot Q=H_0-{\chi\over 2} Q\cdot Q,
\end{equation}
where $H_k$ is the 3D-HO  Hamiltonian for the $k^{th}$ particle, $\chi$ is the strength of the $Q\cdot Q$ interaction and if k and j are two distinct nucleons one has 
\begin{equation}\label{QQ}
Q\cdot Q=\sum_{k,j,m}(-1)^mq_{mk}q_{-mj}=\sum_m(-1)^mQ_mQ_{-m}.
\end{equation}

 The most general equation for the trace of the quadrupole interaction has been derived in the symplectic model (eq.(5.4) of ref. \unscite{Rowe}). If only the valence shell is of interest, with $L=0$ for the $0^+$ energy, one may consider
\begin{equation}\label{Casimir}
Q\cdot Q=<C_{SU(3)}^{(2)}>=4(\lambda ^2+\mu ^2+\lambda \mu+3(\lambda +\mu)),
\end{equation}
where $C^{(2)}_{SU(3)}$ is the second order Casimir operator of SU(3), $(\lambda, \mu)$ are the SU(3) quantum numbers and $<C^{(2)}_{SU(3)}>$ is the eigenvalue of the operator in a given SU(3) irrep.

\section{\normalsize MATRIX REPRESENTATION OF THE U(10) ALGEBRA}

The U(10) algebra of the HO has ten vectors of the type $\ket{n_z,n_x,n_y}$, with total  number of quanta $n=n_z+n_x+n_y=3$. These vectors are the eigenfunctions of the isotropic 3D-HO Hamiltonian \unscite{Cohen}. The generators of SU(3) mentioned in the previous section, can be expressed in matrix representation upon these ten vectors. For convenience the orbitals $\ket{n_z,n_x,n_y}$ are labeled as  $\ket{1}=\ket{3,0,0}$, 
$\ket{2}=\ket{2,1,0}$, $\ket{3}=\ket{2,0,1}$, $\ket{4}=\ket{1,2,0}$, $\ket{5}=\ket{1,1,1}$, $\ket{6}=\ket{1,0,2}$, $\ket{7}=\ket{0,3,0}$, $\ket{8}=\ket{0,2,1}$, $\ket{9}=\ket{0,1,2}$, $\ket{10}=\ket{0,0,3}$. The order of the orbitals is the filling order with nucleons, so as to achieve the highest weight irreducible representation of SU(3) \unscite{lambda-mu}. It is important to note that $\ket{1}$ has $n_\perp=0$, while $\ket{2}$, $\ket{3}$ have $n_\perp=1$ etc.

In ref.\unscite{Lipkin}, chapter 4, eq. (4.5) the three components of angular momentum are defined as $l_{\mu \nu}=x_\mu p_\nu-p_\mu x_\nu=i(a_\mu a_\nu ^\dagger -a_\nu a_\mu ^\dagger)$. Since $[a_\mu,a^\dagger _\nu]=\delta_{\mu \nu}$, it can be derived that $l_z=i(a_y^\dagger a_x-a_x^\dagger a_y)$. The projection of the angular momentum in the matrix representation within the basis $\ket{n_z,n_x,n_y}$ is
\[l_z=
\left[ 
\begin{array}{c@{}c@{}c@{}c}
 \left[\begin{array}{c}
         0  \\
  \end{array}\right]  \\
   & \left[\begin{array}{cc}
                       0 & i\\ 
                       -i & 0
                      \end{array}\right]\\
& & \left[ \begin{array}{ccc}
                                    0 & \sqrt{2}i &0 \\
                                   -\sqrt{2}i &0 & \sqrt{2}i \\
                                   0&-\sqrt{2}i&0
                         \end{array}\right] \\
&&& \left[ \begin{array}{cccc}
                                   0 & \sqrt{3}i & 0 &0 \\
                                   -\sqrt{3}i& 0 & 2i & 0 \\
                                  0 & -2i & 0 & \sqrt{3}i\\
                                   0 & 0 & -\sqrt{3}i & 0
                         \end{array}\right] \\
\end{array}\right]
\]
A diagonalization of each block gives the eigenvalues of the matrix. Specifically the first block has eigenvalue 0, the second $\pm 1$, the third $0, \pm 2$ and the last $\pm 1, \pm 3$. It becomes apparent, that the projection of the angular momentum has values $l_0=\pm 1, \pm 3,...,\pm n_\perp$ if $n_\perp$ is odd and $l_0=0,\pm 2,\pm 4,...,\pm n_\perp$ if $n_\perp$ is even.

From the eqs. (\ref{a}), (\ref{g2}) one can easily derive that
\begin{eqnarray}\label{barq2}
q_2=-{\sqrt{6}\over 2}(n_x-n_y)-{\sqrt{6}\over 2}i(a_x^\dagger a_y+a_y^\dagger a_x),\qquad\qquad
q_{-2}=-{\sqrt{6}\over 2}(n_x-n_y)+{\sqrt{6}\over 2}i(a_x^\dagger a_y+a_y^\dagger a_x).
\end{eqnarray}
From matrix representation and multiplication one gets the matrix of the $\bar q_2=q_{-2}q_{2}+q_2q_{-2}$ interaction
\[q_{-2}q_{2}+q_2q_{-2}=
\left[ 
\begin{array}{c@{}c@{}c@{}c}
 \left[\begin{array}{c}
         0  \\
  \end{array}\right]  \\
   & \left[\begin{array}{cc}
                       6 & 0\\ 
                       0 & 6
                      \end{array}\right]\\
& & \left[ \begin{array}{ccc}
                                   18 & 0 &6 \\
                                   0&12 & 0 \\
                                   6&0&18
                         \end{array}\right] \\
&&& \left[ \begin{array}{cccc}
                                   36 & 0 & 6\sqrt{3} &0 \\
                                   0& 24 & 0 & 6\sqrt{3} \\
                                  6\sqrt{3} & 0 & 24 & 0\\
                                   0 & 6\sqrt{3} & 0 & 36
                         \end{array}\right] \\
\end{array}\right]
\]

This matrix is again organised into blocks containing orbitals with the same $n_\perp$. But since not all blocks are diagonal, an additional quantum number is hidden. This quantum number is the projection of the angular momentum $l_0$. Indeed the commutator is $[q_{\pm 2}q_{\mp 2}, l_0]=0$. Each block has specific values of $n_\perp,l_0$
\[
\left[ 
\begin{array}{c@{}c@{}c@{}c}
 \left[\begin{array}{c}
         n_\perp=0,l_0=0  \\
  \end{array}\right]  \\
   & \left[\begin{array}{cc}
                       n_\perp=1,  & \\ 
                        l_0=\pm 1& 
                      \end{array}\right]\\
& & \left[ \begin{array}{ccc}
                                   n_\perp=2,  &  & \\
                                  l_0=0,\pm 2 & &  \\
                                   &&
                         \end{array}\right] \\
&&& \left[ \begin{array}{cccc}
                                   n_\perp=3,  &  &  & \\
                                   l_0=\pm 1, \pm3&  &  &  \\
                                   &  &  & \\
                                    &  &  & 
                         \end{array}\right] \\
\end{array}\right]
\]
Finally the multiplets of each block are $\ket{n_\perp, l_0}$. 

Possibly an algebraic structure is the cause of this. The answer has been given by J.P. Elliott in section 2 of ref. \unscite{Elliottb}. The same problem is being explained in section 3.6 of ref. \unscite{Harvey}. The algebra SU(3) is decomposed into two subalgebras SU(2) and U(1). The first represents rotations in the $x$-$y$ plane, while the latter involves the $z$ axis. The SU(2) algebra has the following generators $u_{\pm}=\mp \left({1\over 2\sqrt{3}}\right)q_{\pm 2}$, $u_0={1\over 2}l_0$, which satisfy the commutation relations $[u_{\pm},u_0]=\mp u_{\pm}$, $[u_+,u_-]=-u_0$. The U(1) is characterised by the operator $q_0$, which has eigenvalues $\epsilon$ in the Elliott paper \unscite{Elliottb}, with $\epsilon=2n_z-n_\perp.$ The block diagonal form of the matrices $q_{\pm 2}, l_0$ appears due to the SU(2) structure. The SU(2) multiplets are characterised by a quantum number of $j$-type (which is not necessarily the angular momentum) and one of 
$m$-type (needless to say that this is not always the projection of the angular momentum). In the present algebraic structure the $m$-type quantum number is the eigenvalue of $u_0$ and the $j$-type, let it be named $\Lambda_E$ (as in ref. \unscite{Elliottb}, is $\Lambda$), with $\Lambda_E={1\over 2}(n_x+n_y)={1\over 2}n_\perp$. Since $u_0={l_0\over 2}$ and the possible values of $u_0$ are $-\Lambda_E$, $-\Lambda_E +1$, \dots,
$\Lambda_E$ obeying the angular momentum algebra, obviously $l_0=-2\Lambda_E$, $-2\Lambda_E +2$, \dots,$2\Lambda_E$. Therefore for the total number of quanta $n=3$, the possible $n_\perp$ values are $0$, 1, 2, 3. For each one of these,  $\Lambda_E$ takes the values 0, ${1\over 2}$, 1, ${3\over 2}$. If $\Lambda_E =0$ then $l_0=0$ and this reflects to the first block of the matrices $q_{-2}q_2+q_2q_{-2}, l_0$. If $\Lambda_E ={1\over 2}$,  then $l_0=\pm 1$, so this is the second block,  and so on.

\section{\normalsize ACTION Of $\bf{q_2}$ ON NILSSON ORBITALS}

The asymptotic wave functions of Nilsson orbitals are labeled usually as $K[Nn_z\Lambda]$, but there is an alternative notation $\ket{n_zrs\Sigma}$, which is explained very well in section 8.7 of ref. \unscite{book}. A brief repetition of the formalism is necessary.

The Nilsson problem is solved in the cylindrical coordinate system. The quanta on the $x$-$y$ plane are being created by the $R^\dagger$, $S^\dagger$ operators. These operators are exactly the same as the $a_+^\dagger, a_-^\dagger$ respectively, i.e., $R=a_+$, $R^\dagger=a_+^\dagger$, $S=a_-$, $S^\dagger=a_-^\dagger$. The following relations are valid
\begin{eqnarray}
r={N-n_z+\Lambda\over 2},\qquad \qquad s={N-n_z-\Lambda\over 2}, \qquad \qquad \Sigma=K-\Lambda.
\end{eqnarray}
where $R^\dagger R\ket{r}=r\ket{r}$ and $S^\dagger S\ket{s}=s\ket{s}$. The actions $R\ket{r}=\sqrt{r}\ket{r-1}$, $R^\dagger \ket{r}=\sqrt{r+1}\ket{r+1}$, $S\ket{s}=\sqrt{s}\ket{s-1}$, $S^\dagger \ket{s}=\sqrt{s+1}\ket{s+1}$, will be useful. With these tools one can transform the Nilsson orbitals between the two notations $K[Nn_z\Lambda]\rightarrow \ket{n_zrs\Sigma}$. 

The $q_{\pm 2}$ operators with the Nilsson symbols are $q_2=-\sqrt{6}R^\dagger S$, $q_{-2}=-\sqrt{6}S^\dagger R$. Thus the $\bar q_2$ interaction is
$\bar q_2=q_2q_{-2}+q_{-2}q_2=6R^\dagger SS^\dagger R+6S^\dagger RR^\dagger S$. Using that $[R,R^\dagger]=[S,S^\dagger]=1$, the $\bar q_2$ action on Nilsson orbitals becomes
\begin{equation}\label{q2Nilsson}
\bar q_2\ket{n_zrs\Sigma}=6(r+s+2rs)\ket{n_zrs\Sigma}=6(n_\perp +2rs)\ket{n_zrs\Sigma}.
\end{equation}

Now we can calculate the matrix representation of the $\bar q_2$ interaction for two sets of Nilsson orbitals. One set will include the Nilsson orbitals in the shell 20-40, while the other will include the Nilsson orbitals in the shell 28-48. It will be seen that the two matrices are totally identical and diagonal.

The first calculation can be done for the shell  20-40, which includes  the orbitals $\ket{n_zrs\Sigma}$: 
$\ket{300+}$, $\ket{210+}$, $\ket{210-}$, $\ket{120+}$, $\ket{111+}$, $\ket{120-}$, $\ket{030+}$,
$\ket{021+}$, $\ket{021-}$, $\ket{030-}$. This ordering resembles the highest weight order. The same matrix can be created within the shell 28-48. The Nilsson orbitals which are included in this shell are: 
$\ket{400+}$, $\ket{310+}$, $\ket{210-}$, $\ket{220+}$, $\ket{111+}$, $\ket{120-}$, $\ket{130+}$, 
$\ket{021+}$, $\ket{021-}$, $\ket{030-}$. The only difference between the two sets is that 4 out of 10 orbitals of the second set have one more quantum in the $z$ axis, namely
 $\ket{300+}\rightarrow \ket{400+}$, $\ket{210+}\rightarrow \ket{310+}$, $\ket{120+}\rightarrow \ket{220+}$, $\ket{030+}\rightarrow \ket{130+}$. But since the two sets have the same values of $r$ and $s$, they will have the same matrix for the $\bar q_2$ interaction, which for both sets is
\begin{equation}
\bar q_2=q_2q_{-2}+q_{-2}q_2=\left( {\begin{array}{*{20}c}
0&0&0&0&0&0&0&0&0&0\\
0&6&0&0&0&0&0&0&0&0\\
0&0&6&0&0&0&0&0&0&0\\
0&0&0&12&0&0&0&0&0&0\\
0&0&0&0&24&0&0&0&0&0\\
0&0&0&0&0&12&0&0&0&0\\
0&0&0&0&0&0&18&0&0&0\\
0&0&0&0&0&0&0&42&0&0\\
0&0&0&0&0&0&0&0&42&0\\
0&0&0&0&0&0&0&0&0&18
 \end{array} } \right).
\end{equation}

Obviously the $m=2$ component of the quadrupole tensor has the same magnitude in both shells. One could say that part of the quadrupole-quadrupole interaction is incorporated into both shells, due to their identical structure in the $x$-$y$ plane. One shell is defined by the magic numbers of the 3D-HO, which are 2 ,8,  20, 40, 70, 112, \dots, while the other one corresponds to the proxy-SU(3) magic numbers \unscite{calculation}, \unscite{po}. For proton or neutron number greater than 28 the proxy-SU(3) shells close at 48, 80, 124,..., 
which are close to the nuclear magic numbers 50, 82, 126, \dots. In Table \ref{Proxy}, proxy-SU(3) shells are suggested for proton (neutron) numbers less than 50.

\begin{table}
\caption{Proxy-SU(3) magic numbers below 50. The proxy-SU(3) approximation replaces the intruder parity Nilsson orbitals by their $\Delta K[\Delta N\Delta n_z\Delta \Lambda]=0[110]$ counterparts \unscite{calculation}, i.e. $7/2[413]$ is replaced by $7/2[303]$. These counterparts are identical in the 
$x$-$y$ plane, but they differ by one quantum in the $z$ axis. Thus the SU(2) substructure is identical for the orbitals included among the proxy-SU(3) and the HO magic numbers, which have the same $U({(n+1)(n+2)\over 2})$ algebra. }\label{Proxy}
\begin{center}
\begin{tabular}{|c c c c c |}
\hline
shell model & $K[Nn_z\Lambda]$ & 0[110] counterparts & algebra & magic numbers\\
\hline
$1g_{9/2}$ & 9/2[404] & X & & 50\\
& 7/2[413] & 7/2[303] & &48\\
& 5/2[422]& 5/2[312] & & \\
& 3/2[431] & 3/2[321] & & \\
& 1/2[440] & 1/2[330]&&\\
$2p_{1/2}$ & 1/2[301]& & U(10)&\\
$1f_{5/2}$& 5/2[303] & & & \\
&3/2[301] & & & \\
 & 1/2[310]& & &\\
$2p_{3/2}$& 3/2[312]& & &\\
& 1/2[321]& &&\\
\hline
$1f_{7/2}$& 7/2[303]& X & & 28\\
& 5/2[312] & 5/2[202]& & 26\\
&3/2[321]& 3/2[211]& & \\
&1/2[330]& 1/2[220]&&\\
$1d_{3/2}$& 3/2[202]& & U(6)& \\
& 1/2[200]& & &\\
$2s_{1/2}$&  1/2[211]& &&\\
\hline
$1d_{5/2}$& 5/2[202]& X &&14\\
& 3/2[211]& 3/2[101]& U(3)& 12\\
& 1/2[220]& 1/2[110]& &\\
$1p_{1/2}$& 1/2[101]& & & \\
\hline
$1p_{3/2}$& 3/2[101]& X & &6\\
& 1/2[110]& 1/2[000] & U(1)& 4 \\
\hline
$1s_{1/2}$& 1/2[000]& & U(1) &2\\
\hline
\end{tabular}
\end{center}
\end{table}

\section{\normalsize SHAPE COEXISTENCE}

The proxy-SU(3) shells are 0-2, 2-4, 6-12, 14-26, 28-48, 50-80, 82-124, \dots. The corresponding algebras are respectively U(1), U(1), U(3), U(6), U(10), U(15), U(21),... On the contrary, the 3D-HO magic numbers are 2, 8, 20, 40, 70, 112, \dots, defining shells with algebras U(1), U(3), U(6), U(10), U(15), U(21), \dots. The quantum numbers ($\lambda, \mu$) of the highest weight irreps for the algebras U(6), U(10), U(15), U(21) can be found in Table I of ref. \unscite{po}. U(1) has $(\lambda, \mu)=(0,0)$, while the highest weight irreps of U(3) are $(\lambda, \mu)=(2,0),(0,2),(0,0)$ for 2, 4, and 6 valence particles respectively. Thus for each set of magic numbers one can calculate the eigenvalues of the Casimir operator of SU(3), as defined in eq. (\ref{Casimir}). The resulting plot is Fig.(\ref{Casimirs}). 

The Casimir operator is proportional to the quantities $\beta^2$, $B(E2, 2_1^+\rightarrow 0_1^+)$. Thus it measures the deformation \unscite{Bible}. In the set of the 3D-HO magic numbers these eigenvalues start to become smaller, compared to the ones obtained for the proxy-SU(3) calculation, at the proton (neutron) numbers 8, 18, 34, 60, 96, 146, 210. From these numbers up to the neighboring shell closure of the 3D-HO, i.e., 8, 20, 40, 70, 112, 168, 240, the two sets of magic numbers compete against each other and the result is shape coexistence. So in proton (neutron) numbers within the regions {\bf 8, 18-20, 34-40, 60-70, 96-112, 146-168, 210-240} one may expect that the experimental data will show the appearance of two $0^+$ energy bands. The ground state $0^+$ is usually the outcome of the proxy-SU(3) magic numbers, while in the excited $0^+$ the 3D-HO is somehow involved. At Fig. 8 of ref. \unscite{coexistence} an approximate map of shape coexistence manifestation is presented. All areas have protons or neutrons among the previously proposed numbers. Furthermore, shape coexistence appears from 96 to 110 neutrons in the Hg isotopes and from 60 to 70 neutrons in the Sn isotopes (Fig. 10 of ref. \unscite{coexistence} and Fig. 3.10 of \unscite{Wood}). Clearly shape coexistence ends in the 3D-HO shell closure.

\begin{figure}\label{Casimirs}
\begin{center}
\caption{The values of $\lambda ^2+\mu ^2+\lambda \mu+3(\lambda+\mu)$ for protons (neutrons) for the two sets of magic numbers (proxy-SU(3) and 3D-HO). In most experimental cases shape coexistence begins when the Casimir, as calculated by the 3D-HO magic numbers, becomes less than the Casimir of proxy-SU(3) magic numbers. The phenomenon ends at the 3D-HO shell closure.}\label{Casimirs}
\includegraphics[scale=0.6]{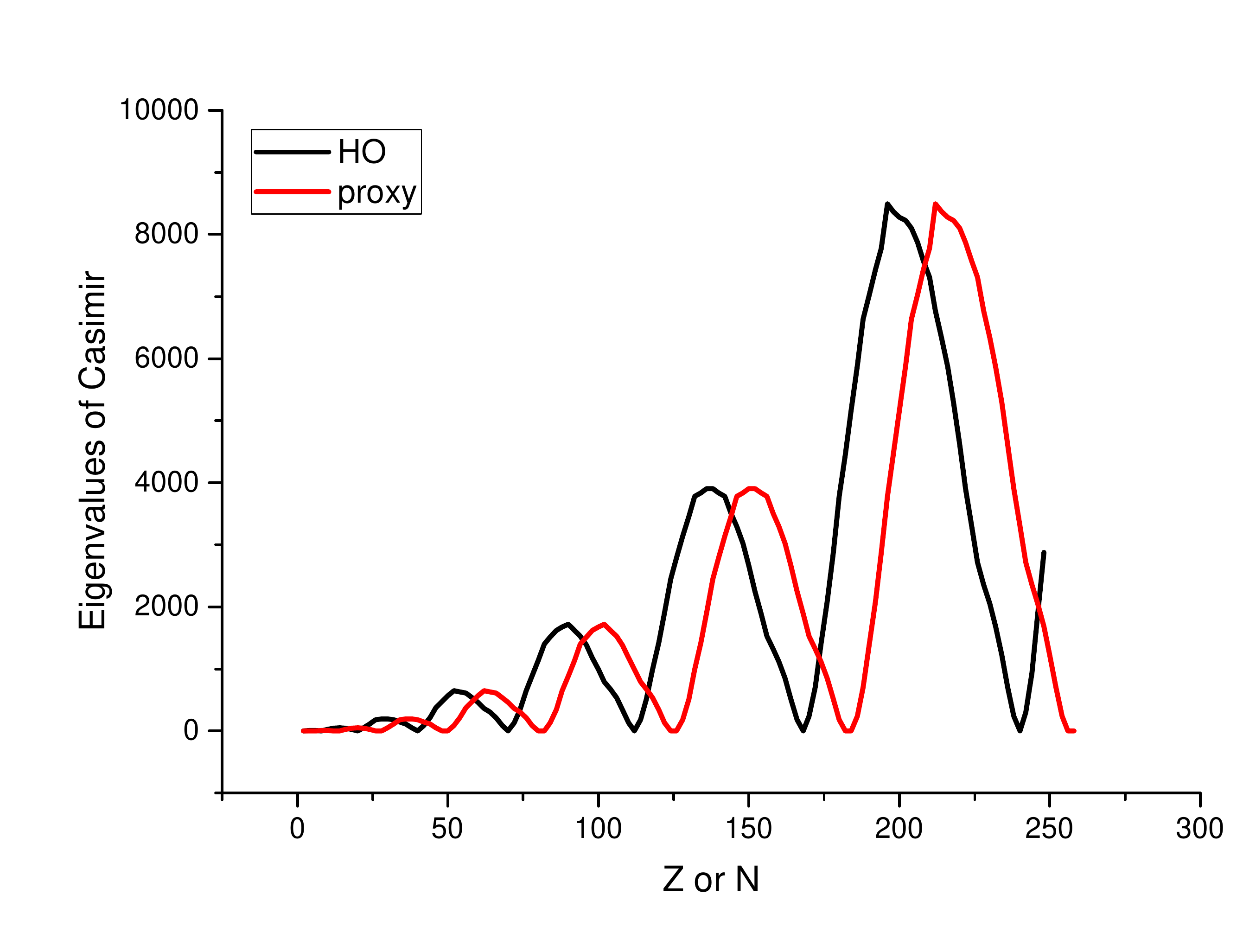}
\end{center}
\end{figure}

\end{document}